\documentstyle[epsf,epsfig,rotate]{aa}
\onecolumn

\begin{document}

\thesaurus{02(02.01.2; 02.13.2; 08.09.2 AA Tau; 08.13.1; 08.16.5)}

\title{The response of an accretion disc to an inclined dipole with
application to AA~Tau}

\author{Caroline Terquem \inst{1,2,3}
        \and John C. B. Papaloizou \inst{4}}

\institute{Institut d'Atrophysique de Paris, 98 bis Boulevard Arago,
       75014 Paris, France -- terquem@iap.fr \and Universit\'e Denis
       Diderot--Paris VII, 2 Place Jussieu, 75251 Paris Cedex 5,
       France \and Laboratoire d'Astrophysique, Observatoire de
       Grenoble, BP 53, 38041 Grenoble Cedex 9, France \and Astronomy
       Unit, School of Mathematical Sciences, Queen Mary \& Westfield
       College, Mile End Road, London E1 4NS, UK --
       J.C.B.Papaloizou@maths.qmw.ac.uk}

\offprints{C. Terquem}
\date{Received / Accepted: 07 June 2000}

\titlerunning{The response of an accretion disc to an inclined dipole}
\authorrunning{Terquem \& Papaloizou}

\maketitle


\begin{abstract}

We compute the warping of a disc induced by an inclined dipole.  We
consider a magnetised star surrounded by a thin Keplerian diamagnetic
disc with an inner edge that corotates with the star.  We suppose the
stellar field is a dipole with an axis that is slightly misaligned
with the stellar rotation axis.  The rotation axes of the disc
material orbiting at large distances from the star and that of the
star are supposed to coincide.  The misalignment of the magnetic and
rotation axes results in the magnetic pressure not being the same on
the upper and lower surfaces of the disc.  The resultant net vertical
force produces a warp which appears stationary in a frame corotating
with the star.  We find that, if viscosity is large enough ($\alpha
\sim 0.01$--0.1) to damp bending waves as they propagate away, a
smoothly varying warp of the inner region of the disc is produced.
The amplitude of the warp can easily be on the order of ten percent of
the disc inner radius for reasonably small misalignment angles (less
than 30 degrees).  Viscous damping also introduces a phase shift
between the warp and the forcing torque, which results in the
locations of maximum elevation above the disc forming a trailing
spiral pattern.  We apply these results to recent observations of
AA~Tau, and show that the variability of its light curve, which occurs
with a period comparable to the expected stellar rotation period,
could be due to obscuration produced by a warp configuration of the
type we obtain.


\end{abstract}

\section{Introduction}

Objects accreting material through an accretion disc very commonly
contain a significant magnetic field.  This is the case for accreting
white dwarfs in cataclysmic variables, some X--ray binary pulsars and
at least some classical T~Tauri stars (CCTS).

It was first suggested by Bertout et al. (\cite{Bertout}), as a result
of the detection of bright stellar spots, that CTTS may accrete along
stellar magnetic field lines.  This picture has been further supported
by a wide array of observational evidence (see Najita et
al. \cite{Najita} and references therein), including spectroscopic
indications of infalling material onto the stellar surface (Edwards et
al. \cite{Edwards2}; Hartmann et al. \cite{Hartmann}) and the low spin
rate of CCTS (Bouvier et al. \cite{Bouvier1}; Edwards et
al. \cite{Edwards1}).

Since T~Tauri stars have a large convective envelope, it is likely
that at least part of their magnetic field is generated through a
dynamo process.  However, there may also be a fossil component
originating from the molecular cloud out of which the star formed
(Tayler \cite{Tayler}).  Recent Zeeman measurements indicate
relatively strong field strength at the surface of T~Tauri stars, on
the order of one kilogauss (Guenther et al. \cite{Guenther};
Johns-Krull et al. \cite{Johns}).  It is not known what the structure
of the field is.  At some distance from the star the dipolar component
probably dominates, but whether this is the case in the magnetosphere
is not clear.  However, observations cannot rule out such a coherent
field structure (Montmerle et al. \cite{Montmerle}), and numerical
simulations of nonlinear stellar dynamos indicate that a steady dipole
mode is the most easily excited one (Brandenburg et al. \cite{brand1}).

Interaction between the stellar magnetic field and the accretion disc
has very important consequences for the disc structure, the accretion
process (see Ghosh \& Lamb \cite{Ghosh91} and references therein) and
the evolution of the stellar rotation (K\"onigl \cite{Konigl}).  In
particular, the disc is truncated by the magnetic torque, so that it
does not extend down to the stellar surface (Gosh \& Lamb
\cite{Ghosh79}).  The location of the disc inner radius is determined
by the condition that magnetic and viscous torques balance.  For CTTS,
the radius of the inner cavity is believed to be a few stellar radii
(see, e.g., Wang \cite{Wang}).

So far, there are only a few numerical simulations of disc--stellar
magnetic field interactions (Hayashi et al. \cite{Hayashi}; Miller \&
Stone \cite{Miller}; Goodson et al. \cite{Goodson}; Kudoh et
al. \cite{Kudoh}).  They all show the disc--magnetosphere interaction
to be complex and sensitive to initial and boundary conditions.  At
this stage, it is not clear what final form a full theoretical model is
likely to take.  However, analytical or semi--analytical simplified
models can still be valuable in pointing out some important processes
that may arise in these systems, and the goal of this paper is to
describe one of these processes.

We note that the magnetic axis and the rotation axis of the disc at
large distances from the star may not be aligned, although often, for
simplicity, they are assumed to coincide.  We assume here that the
stellar rotation axis and the disc rotation axis at large distances
coincide.  Misalignment would then occur if, for instance, the star
were to generate a dipole field with magnetic axis misaligned with its
spin axis (like in the case of the Earth).  It is possible that the
interaction with the disc would lead to some evolution of the
misalignment angle, but the details are likely to depend on the
processes which generate the field.  In any case, when such a
misalignment is present, the magnetic pressure is not the same on the
upper and lower surfaces of the disc.  This mismatch generates a net
vertical force which excites bending waves and warps inner parts of
the disc (Aly \cite{Aly}).

Bending instabilities in a disc subject to a stellar magnetic dipole
have been investigated by Agapitou et al. (\cite{Agapitou}, hereafter
APT).  APT calculated the global bending modes of a disc permeated by
both an internally produced poloidal magnetic field and an external
dipole field with axis aligned with the disc rotation axis (in this
case no warp is induced by the dipole configuration, but free bending
modes can be excited by a perturbation which takes the disc out of its
equilibrium plane).  They found that instability could occur if the
magnetic and centrifugal forces were comparable in some region of the
disc.  They pointed out that such instabilities may result in the
periodic variability observed in the light curve of many CTTS.
                              
Lai (\cite{Lai}) studied the warping of a disc induced by an inclined
dipole.  He calculated the magnetic torque exerted by an inclined
dipole on a disc, and studied the stability against vertical
displacements of a disc subject to such a torque.  In the terms of the
APT analysis, he studied the stability of low frequency (as measured
in an inertial frame) bending modes corresponding to the modified tilt
mode as discussed in APT.  We note that, when considering the
structure of the disc subject to the inclined dipole, he did not take
into account the effects of the distortion of the disc itself on its
response, which can have important consequences on the dynamics
through wave propagation.  But he added the effects of a toroidal
field, assumed to be generated by winding up a penetrating vertical
field, on the magnetic pressure determining the vertical force on the
disc. This contribution is phase shifted with respect to the other
contributions and may thus (if not counteracted) cause the modified
tilt mode to become unstable, resulting in spontaneous warping.  To
decide whether this mode can be destabilised requires detailed
consideration of the effects of wave propagation and viscosity.  We
comment that under some conditions warps diffuse away on a timescale
much shorter than the viscous timescale (Papaloizou \& Pringle
\cite{Pap2}) or propagate away with a velocity on the order of the
sound speed (Papaloizou \& Lin \cite{Pap1}) resulting in
stabilisation.
    
In this paper, we calculate the structure of a thin Keplerian disc
subject to an inclined dipole, taking into account the effects of the
distortion of the disc itself on its response, i.e. the full dynamics
of the system.  For simplicity, we suppose that the disc is
diamagnetic, so that it is not permeated by the external stellar
field.  In principle, the calculations presented here could be
extended to more general cases.  However, if the disc were not
diamagnetic, wrapping of field lines would probably become important,
leading to the possible disruption of the magnetosphere (see, e.g.,
Mikic \& Linker \cite{Mikic}).  Also we do not address here the
physical processes of accretion or plasma entry into the stellar
magnetosphere.  We note that because the warp induced in the inner
disc appears steady in a frame rotating with the star, any resulting
variability would have the same period as that of the star.

We comment that the generation of spontaneous warping does not apply
to the calculations we present here, since we study a response which
is forced by the inclined dipole and has a pattern speed equal to the
rotation rate of the star. Thus, in contrast to the considerations of
Lai (\cite{Lai}), it is not a modified tilt mode.

This work has been motivated by a recent study of Bouvier et
al. (\cite{Bouvier2}) who report that the light curve of the CTTS
AA~Tau displays photometric, spectroscopic and polarimetric variations
on timescales from a few hours to several weeks.  The most striking
feature of this light curve is a photometric variability with a period
comparable to the expected rotation period of the star.  This has been
interpreted by Bouvier et al. (\cite{Bouvier2}) as being due to the
occultation of the star by a warp of the inner disc (the system is
observed almost edge--on).  The authors speculated that the warp could
be produced by an inclined dipole.  We note that Bouvier et
al. (\cite{Bouvier2}) did not consider AA~Tau as being a special case
as far as its properties are concerned.  They pointed out that only
its light curve is unusual, and they interpreted it as being due to
the fact that the system is seen almost edge--on.  In other words,
warping of the inner parts of CTTS discs would not be uncommon, but it
could be seen only for particular viewing angles.

The plan of the paper is as follows: We begin by considering an
equilibrium configuration where the axis of the dipole and the
rotation axis of the disc are aligned.  This is described in
\S~\ref{sec:equil}.  We then perturb this equilibrium by slightly
inclining the dipole. In \S~\ref{sec:pert} we calculate the resulting
perturbed magnetic field and derive the integro--differential equation
which has to be solved for the disc vertical displacement.  This
equation is solved in the WKB approximation, which is valid when the
wavelength of the bending waves excited in response to the
perturbation is small compared to the disc radius.  To allow these
waves to damp as they propagate away (and therefore to damp very small
wavelength oscillations, which would be unphysical), we include the
effects of viscous damping in the integro--differential equation.  In
\S~\ref{sec:results} we solve this equation numerically and present
the results for two different magnetic field equilibria (derived by
Aly \cite{Aly} and Low \cite{Low}).  In both cases we find that, if
the viscosity is large enough to damp the waves as they propagate
away, a smooth warp configuration of the disc inner parts can exist.
The elevation above the disc equilibrium plane (i.e. the stellar
equatorial plane) can easily be on the order of ten percent of the
disc inner radius for reasonably small misalignment angles (less than
30 degrees).  If the viscosity is too small to damp the waves
efficiently, the disc inner parts may be disrupted, and it is likely
that the evolution is then highly time--dependent.  In
\S~\ref{sec:discussion} we apply these results to the case of AA~Tau,
and show that the variability of its light curve, which occurs with a
period expected to be the stellar rotation period, can plausibly be
explained by a warp configuration of the type we obtain.

\section{Disc and aligned dipole}

\label{sec:equil}

We begin by considering a thin disc configuration such that the gas
orbits a central rotating star with a dipole field and where the
magnetic axis and rotation axes are all aligned.  In this situation,
the configuration is axisymmetric.  For convenience, we work in a
frame corotating with the central star with angular velocity $\omega.$
The disc is truncated in its inner parts by the magnetic torque and we
will suppose that the inner edge corotates with the star.

\subsection{Magnetic field}
\label{sec:mfield}

The dipole field ${\mathbf B}_{ext}$ due to the central star induces
azimuthal currents in the conducting disc which in turn generate an
additional poloidal magnetic field ${\mathbf B}_d$.

Here, as in APT, we shall assume that the field external to the disc
and central star can be approximated as curl free, i.e. assumed to be
a vacuum field (the currents external to the disc and star are at
infinity).

The total axisymmetric poloidal magnetic field, ${\mathbf B}_{ext} +
{\mathbf B}_d$, is noted ${\mathbf B}= (B_r,0,B_z)$, where we use
cylindrical polar coordinates $(r,\varphi, z)$.  The associated
Cartesian coordinates are $(x,y,z)$.  The flux function $\psi$ is such
that:

\begin{equation} 
B_r= {-1 \over r}{\partial \psi \over \partial z} \; {\rm and}\ B_z =
{1\over r}{\partial\psi\over \partial r}.
\label{psi}
\end{equation} 

\noindent For the stellar dipole field:

\begin{equation}
\psi_{ext} = -{\mu_d r^2 \over( r^2 +z^2 )^{3/2 }},
\label{psid}
\end{equation} 

\noindent where the magnitude of the stellar magnetic dipole moment is
$\mu_d=B_{\ast} R_{\ast}^3$, with $B_{\ast}$ and $R_{\ast}$ being the
stellar magnetic field and radius, respectively.

\noindent For the general axisymmetric poloidal field, the associated
current density is ${\mathbf j}=(0,j_{\varphi},0).$ For an
infinitesimally thin disc, as we consider here, we define the
vertically integrated azimuthal component of the current density, $J,$
such that:

\begin{displaymath} 
J=\int^{\infty}_{-\infty} j_{\varphi}{\rm d}z.
\end{displaymath} 

\noindent By integrating the azimuthal component of Amp\`ere's law
through the disc, we obtain:

\begin{equation}
B_r^{+}= \mu_0J/2,
\label{br}
\end{equation} 

\noindent where $B_r^{+}$ denotes the radial component of the magnetic
field just outside the upper surface of the disc.  Throughout this
paper we use MKSA units, and $\mu_0$ denotes the permeability of the
vacuum.  $B_r$ is antisymmetric with respect to reflection in the disc
mid--plane so that its value just outside the lower surface of the
disc is $B_r^{-}=-B_r^{+}.$

\subsection{Force balance}

\label{sec:forceb}

Neglecting pressure forces as in APT, the vertical integration of the
radial component of the momentum equation yields the condition for
radial equilibrium as:

\begin{equation} 
\Sigma {\partial \Phi \over \partial r}=\Sigma r \Omega^2 + J B_z,
\label{EQ}
\end{equation} 

\noindent where $\Sigma$ is the surface density, $\Omega$ is the disc
angular velocity measured in an inertial frame, and $\Phi$ is the
gravitational potential here taken to be due to a central point mass,
$M_{\ast}$, such that:

\begin{displaymath} 
\Phi =-{GM_{\ast} \over \sqrt{r^2+z^2}}.
\end{displaymath} 

\noindent If $B_z \ne 0,$ in some regions of the disc a variety of
configurations are possible (see APT).  These include cases where
inner field lines cross the disc and join to the central star and thus
may be assumed to corotate with it.  Outer field lines may be open in
the case of an infinite disc or a finite disc with appropriate
boundary conditions.

In this paper, for simplicity, we shall perform calculations for the
special case where the field is excluded from the disc.  This
situation arises when the disc is perfectly conducting.  Then only
surface currents flow in the disc and they screen the stellar dipole
field from the disc interior, i.e. they produce an additional field
${\mathbf B}_d$ which cancels ${\mathbf B}_{ext}$ in the disc
interior.  We note that, since the vertical component of the field is
continuous at the disc surface, $B_z$ just outside the disc is zero in
this case.  The same does not apply for $B_r$, and in general $B^+_r$
and $B^-_r$ are non zero.

We remark that, since there is no Lorentz force acting in the disc
when $B_z=0$, the angular velocity given by equation~(\ref{EQ}) is
Keplerian.

\section{Disc and slightly misaligned dipole}
\label{sec:pert}

We suppose that the system is perturbed from the axisymmetric
equilibrium state described above by introducing a slight
misalignment between the magnetic axis of the central dipole and the
rotation axis of the disc and star (the $z$ axis).  This produces a
non axisymmetric response in the disc that can be described using
linear perturbation theory.

The main features of this response is that it takes the form of a
warping of the disc, as indicated by Aly (\cite{Aly}), together with
the additional feature of the excitation of bending waves.  The
response is naturally largest in the inner parts of the disc where the
magnetic field is strongest.  As we indicate by considering specific
examples in \S~\ref{sec:results}, it can take the form of a steeply
changing inclination of the inner disc orbits which can make it appear
to have an inner wall.

To calculate the geometry of the disc (that is its elevation above its
equilibrium plane), we extend the analysis of APT.  To compute the
free bending modes of the disc, APT solved an eigenvalue problem where
the eigenvalues were the mode frequencies and the eigenfunctions their
amplitudes.  Here we are interested in a response with frequency equal
to that of the forcing term.  Since the dipole is anchored in the
star, this is the angular velocity of the star.  The amplitude of the
mode (its spatial dependence) can be found from the mode equation of
APT with a specified frequency and the addition of the forcing term.
However, a different equilibrium field is adopted, since here, in
contrast to APT, there is no internally produced field in the disc.

\subsection{Perturbed magnetic field}
\label{sec:bpert}

To calculate the response we suppose the central dipole moment is
rotated in the $(x,z)$ plane through a small angle $\delta$ being the
inclination to the $z$ axis.  The dipole moment is then given by:

\begin{equation}
\mbox{\boldmath $\mu$}_d = (\mu_d \delta, 0, \mu_d ).
\label{psidm}
\end{equation}    

\noindent This contributes to a potential:

\begin{equation}
\Phi'_{{\rm M},ext} =
\frac{ \mbox{\boldmath $\mu$}_d \cdot {\mathbf r}}{r^3} ,
\label{phiext1}
\end{equation}

\noindent where ${\mathbf r}$ denotes the position vector, which
produces a radial external magnetic field perturbation
$B'_{r,ext}=\partial \Phi'_{{\rm M},ext} / \partial r$.  Just outside
the disc surfaces, this is given by the real part of:

\begin{equation}
B'^+_{r, ext} = B'^-_{r, ext} 
 = -{ 2 \mu_d \delta \exp(i\varphi) \over r^3}.
\label{bext}
\end{equation}  

Thus the problem reduces to the calculation of the disc response to a
field perturbation with azimuthal mode number $m=1.$ This, when acting
with the unperturbed azimuthal current, produces a vertical Lorentz
force which tends to warp the disc.  In other terms, when the dipole
is misaligned, the magnetic pressure force is not the same on the
upper and lower surfaces of the disc.  This mismatch generates a net
vertical pressure force which tends to warp the disc.

We thus introduce the Lagrangian displacement $\mbox{\boldmath $\xi$}$
which, in a razor--thin disc approximation, has the form:

\begin{displaymath}
\mbox{\boldmath $\xi$} = (0,0,\xi_z).
\end{displaymath}

\noindent The only non-negligible component is the vertical one (APT),
and $\xi_z$ represents the elevation above the disc equilibrium plane.
In keeping with the form of the external magnetic field perturbation,
we make all the perturbed quantities complex by taking their
$\varphi$-dependence to be through a factor $\exp( i \varphi)$, which
henceforth will be dropped.  The physical perturbations will be
recovered by taking the real part of these complex quantities.  The
Eulerian perturbations of the various quantities are denoted by a
prime.

\noindent The perturbation of the magnetic field interior to the disc,
${\mathbf B}'$, is related to $\mbox{\boldmath $\xi$}$ by the flux
freezing condition:

\begin{equation}
{\mathbf B}' = (B'_r, B'_{\varphi}, B'_z) = \mbox{\boldmath $\nabla$}
\mbox{\boldmath $\times$} (\mbox{\boldmath $\xi$} \mbox{\boldmath
$\times$} {\mathbf B}).
\end{equation}

\noindent The non-zero components of ${\mathbf B}'$ take the form:

\begin{equation}
B'_r = -\xi_z {\partial B_r\over \partial z},\ {\rm and }\  
B'_z= {1 \over r}{\partial (rB_r \xi_z)\over \partial r} .
\label{bpert}
\end{equation}

\noindent Since $B_r$ is antisymmetric with respect to reflection in
the disc mid--plane and $\xi_z$ is independent of $z$ in the thin disc
approximation, $B'_z$ is antisymmetric and $B'_r$ is symmetric.

\noindent As in APT, the vertical component of the perturbed field
must be matched to the vertical component of a perturbed vacuum field
exterior to the disc.  Here this is taken to be a potential field.
Therefore we have:

\begin{equation}
\frac{\partial \Phi'_{\rm M}}{\partial z} = B'^{+}_z, 
\label{derphi}
\end{equation}

\noindent where $B'^{+}_z$ is the value of the vertical field
perturbation just outside the disc surface and $\Phi'_{\rm M}$ is the
magnetic potential associated with the external field perturbation.
To find $\Phi'_{\rm M}$, we first subtract out the contribution of
the external field perturbation arising from the tilted dipole.  At
the disc surface, this is (eq.~[\ref{phiext1}]):

\begin{equation}
\Phi'_{{\rm M},ext}
 = { \mu_d \delta \over r^2},
\label{phiext}
\end{equation}      

\noindent where as above the factor ${\rm exp}(i \varphi)$ has been
dropped.  With this contribution removed, the residual potential has
no singularity outside the disc.  It can be calculated from
equation~(\ref{derphi}) in an analogous manner to finding the
gravitational potential due to a disc surface density distribution,
with $2 \pi G \Sigma$ (where $G$ is the gravitational constant) being
replaced by $B'^+_z$ (see Tagger et al. \cite{Tagger}; Spruit et
al. \cite{Spruit}; APT).  Following this procedure, $\Phi_{\rm M}'$
may be written as the sum of a Poisson integral and $\Phi'_{{\rm
M},ext}$:

\begin{equation} 
\Phi'_{\rm M} = \Phi'_{{\rm M},d} + \Phi'_{{\rm M},ext} ,
\label{phisum}
\end{equation}

\noindent where

\begin{equation} 
\Phi'_{{\rm M},d}= - {1 \over 2 \pi} \int^{R_o}_{R_i} \int^{2 \pi}_0
{ B'^{+}_z(r') \cos \varphi' \; r' {\rm d}r' {\rm d}\varphi' \over
\sqrt{ r'^2 + r^2 - 2rr'\cos(\varphi') + z^2 } },
\label{mpot}
\end{equation}

\noindent where $R_i$ and $R_o$ are the inner and outer radii of the
disc, respectively.  In the above integral, the $\varphi$-dependence
of the perturbed field has been taken into account.  Here again, the
factor $\exp( i \varphi)$, to which $\Phi'_{{\rm M},d}$ is
proportional, has been dropped.

Since the disc is perfectly conducting, the vertical component of the
field given by equation~(\ref{bpert}) is continuous at the disc
surface.  Therefore:

\begin{equation}
B'^{+}_z = {1 \over r}{\partial (rB^{+}_r
\xi_z)\over
 \partial r}.
\label{bpzp}
\end{equation}

\noindent This expression for $B'^+_z$ can be used in the above
integral.

The radial component of the magnetic field just outside the surfaces
of the disc is given by:

\begin{equation}
B'^{+}_r= B'^{-}_r =\left({\partial \Phi'_{\rm M} \over \partial
r}\right)_{z=0} = \left({\partial \Phi'_{{\rm M},d} \over \partial
r}\right)_{z=0} - {2 \mu_d \delta \over r^3} ,
\label{pot}
\end{equation}

\noindent where equations~(\ref{phiext}) and~(\ref{phisum}) have been
used to write the last expression (we have approximated the
derivatives at the disc surfaces by their value at $z=0$ because the
disc is infinitesimally thin).

\subsection{ The disc vertical displacement}

\noindent The vertical component of the perturbed Lorentz force
integrated vertically through the disc is (see APT):

\begin{equation} 
\int^{\infty}_{-\infty} F'_z {\rm d}z = -{2 B_r^{+} B_r'^{+} \over
\mu_0} 
.
\label{force}
\end{equation} 

\noindent The term proportional to $\partial B_z / \partial r$, which
was present in APT, is zero here since $B_z=0$ in the disc interior
and at its surfaces (see \S~\ref{sec:forceb}).

We note that this force is simply the perturbed magnetic pressure
force $P'_m$ vertically integrated through the disc.  The pressure
force vertically integrated through the disc is indeed:

\begin{equation}
P_m = - \frac{ \left( B_r^{+} \right)^2}{2 \mu_0} +
\frac{ \left( B_r^{-} \right)^2}{2 \mu_0} ,
\label{pm}
\end{equation}

\noindent ($B_z=0$ at the disc surfaces), so that:

\begin{equation}
P'_m = - \frac{B^+_r B'^+_r}{\mu_0} + \frac{B^-_r B'^-_r}{\mu_0}
= - \frac{2 B^+_r B'^+_r}{\mu_0},
\label{Ppert}
\end{equation}

\noindent where we have used the symmetries of the field.  We note
that $P'_m$ is non zero because the perturbation induced by the
misaligned dipole has an opposite effect on the upper and lower
surfaces of the disc: it increases the magnitude of the radial
component of the magnetic field on one of the surfaces from $|B^+_r|$
to $|B^+_r|+|B'^+_r|$ while decreasing it on the other surface from
$|B^+_r|$ to $|B^+_r|-|B'^+_r|$.

The vertically integrated $z$-component of the perturbed equation of
motion is:

\begin{equation} 
\Sigma{{\rm D}^2 \xi_z \over {\rm D}t^2}= -\Sigma \left({\partial^2
\Phi \over \partial z^2}\right)_{z=0} \xi_z + \int^{\infty}_{-\infty}
F'_z {\rm d}z
\label{mot}
\end{equation} 

\noindent where ${\rm D}/{\rm D}t$ denotes the convective derivative.

\noindent In the problem considered here, the solution is steady in a
frame rotating with the central star angular velocity $\omega.$ Then:

\begin{equation}
\Sigma{{\rm D}^2 \xi_z \over {\rm D}t^2}= -\Sigma(\omega -\Omega)^2\xi_z.
\label{moto}
\end{equation}           

\noindent Also, for a point mass potential:

\begin{equation}
\left({\partial^2 \Phi \over \partial
z^2}\right)_{z=0}={GM \over r^3}=\Omega_{\rm K}^2, 
\label{gravo}
\end{equation}

\noindent where $\Omega_{\rm K}$ is the Keplerian angular velocity.
Using equations~(\ref{pot}), (\ref{force}), (\ref{moto})
and~(\ref{gravo}), equation~(\ref{mot}) becomes:

\begin{equation}
\left[ - (\omega-\Omega)^2 + \Omega_{\rm K}^2 
\right] \xi_z = - {2B_r^+ \over \mu_0 \Sigma} \left[ \left( {\partial
\Phi'_{{\rm M},d} \over \partial r} \right)_{z=0} - {2 \mu_d \delta
\over r^3} \right] .
\label{mod1}
\end{equation}

\noindent The potential $\Phi'_{{\rm M},d}$ can be expressed in term
of $\xi_z$ and its derivative with respect to $r$ using
eq.~(\ref{mpot}) and~(\ref{bpzp}).  Therefore equation~(\ref{mod1})
gives a linear equation for determining the vertical displacement
$\xi_z$ induced by the tilted external dipole.  The term proportional
to $\mu_d$ in this equation acts like a forcing term for the vertical
displacement and it causes the disc to become warped and may also
excite bending waves.

\noindent As noted above, since there is no Lorentz force acting in
the disc interior, $\Omega$ in equation~(\ref{mod1}) is the Keplerian
angular velocity.

We remark that equation~(\ref{mod1}) is the same as equation~(14) of
APT with $B_z=0$, but with an additional nonhomogeneous term
proportional to $\mu_d$.  As noted above, this is expected as here we
calculate the response which is forced by the nonaligned dipole.  We
remark that this nonhomogeneous term is simply the pressure force due
to the misaligned component of the dipole vertically integrated
through the disc:

\begin{equation}
P'_{m,ext} = - \frac{2 B^+_r B'^+_{r,ext}}{\mu_0} = \frac{2
B^+_r}{\mu_0} \frac{2 \mu_d \delta}{r^3},
\end{equation}

\noindent where we have used equation~(\ref{bext}) (see also eq.~[24]
of Aly \cite{Aly}).

\subsection{WKB waves}

\label{sec:dispersion}

When $\mu_d$ is set to zero in equation~(\ref{mod1}), this equation
has solutions corresponding to free bending waves.  In the local
limit, these can be found by assuming that $\xi_z \propto \exp(ikr),$
where $k$ is the radial wavenumber, assumed to be very large compared
to $1/r$.  The integral in equation~(\ref{mpot}) can be evaluated in
the WKB approximation to give, after using (\ref{bpzp}) (see APT and
references therein):

\begin{equation} 
\Phi'_{\rm M} = \frac{-B'_z}{|k|} = \frac{ - i k B_r^{+} \xi_z}{|k|} .
\end{equation}

\noindent The local dispersion relation derived from
equation~(\ref{mod1}) is then:

\begin{equation} 
(\omega - \Omega_{\rm K})^2= \Omega_{\rm K}^2
+ {2(B_r^{+})^2\over \mu_0 \Sigma}|k|,
\label{fb} 
\end{equation}

\noindent where we have used the fact that the angular velocity is
Keplerian.  We see from equation~(\ref{fb}) that bending waves with
$m=1$ propagate (with $|k|>0$) exterior to the Lindblad resonances
where $(\omega-\Omega_{\rm K}) = \pm \Omega_{\rm K}$.  In this paper,
we are concerned with the situation where the disc is terminated at an
inner boundary $(r=R_i)$ where the local Keplerian rotation rate is
close to corotation with the central star, i.e. $\Omega_{\rm K}(R_i) =
\omega$.  In such a case, only the outer Lindblad resonance (OLR),
where $\Omega_{\rm K} = \omega /2$, will exist within the disc, and it
occurs at a radius $r= 1.59 R_i.$

\noindent Significantly beyond the OLR, the wavenumber associated with
the waves is given by:

\begin{equation}
|k| r \sim {\mu_0  \omega^2 r  \Sigma \over 2 (B_r^{+})^2 } 
\propto \frac{r^3}{\beta}, 
\end{equation}

\noindent where we define $\beta$ such that it would be the ratio of
magnetic to centrifugal forces if we had a vertical field $B_z \sim
B^+_r$ (see \S~\ref{sec:results}).  Although $B_z=0$ here, we use
$\beta$ as a measure of the strength of $B^+_r$.  This is expected to
be at most of order unity at the inner edge of the disc and then to
decrease rapidly outwards, as the magnetic field decreases
(see \S~\ref{sec:results}).

If $\beta$ decreases as $r$ increases, the forcing term proportional
to $\mu_d$ in equation~(\ref{mod1}) will excite bending waves in the
disc that propagate outwards with increasing wavenumber until they are
damped, much as in the case of Saturn's rings (see Goldreich \&
Tremaine \cite{Goldreich}; Shu \cite{Shu}). In some cases, the
wavelength may be so short at the OLR that dissipative processes
prevent waves from being properly launched (see below).

\subsection{Viscous dissipation}

In order to allow the waves to damp as they propagate away from the
OLR, we add an additional viscous force per unit mass to the right
hand side of equation (\ref{mot}) of the form $\nu \partial^2 v'_z /
\partial r^2$, where $v'_z$ is the Eulerian perturbed vertical
velocity and $\nu$ is a kinematic viscosity.  Neglecting the variation
of $\Omega_K,$ which is reasonable for short wavelength disturbances,
this can also be written as $i \nu (\Omega_{\rm K} -\omega) \partial^2
\xi_z / \partial r^2$.  For $\nu$ we use a standard '$\alpha$'
prescription such that $\nu = \alpha H^2 \Omega_{\rm K}$, where $H$ is
the disc semi--thickness and $\alpha$ is a constant (Shakura \&
Sunyaev \cite{Shakura}).

\noindent With the incorporation of the above viscous force, equation
(\ref{mod1}), which gives the forced response of the disc, becomes:

\begin{equation}
\left[ - ( \omega - \Omega)^2 + \Omega_{\rm K}^2 
\right] \xi_z - i \nu ( \Omega_{\rm K} - \omega) {\partial^2 \xi_z
\over \partial r^2} = - { 2 B_r^{+} \over \mu_0 \Sigma } \left[ \left( {
\partial \Phi'_{{\rm M},d} \over \partial r } \right)_{z=0} - {2 \mu_d
\delta \over r^3} \right] .
\label{mod}
\end{equation}    

We now comment briefly on the possible origin of this $\alpha$
viscosity.  So far, the only process which has been shown to initiate
and sustain turbulence in Keplerian discs is the magnetohydrodynamic
Balbus--Hawley instability, and it does lend itself to an $\alpha$
formalism (Balbus \& Hawley \cite{Balbus} and references therein).
For the disc we consider here, which is not permeated by the external
magnetic field, it is difficult to justify the existence of an
$\alpha$ viscosity.  However, in reality, the disc is probably
permeated to at least a small extent by the external field (Ghosh \&
Lamb \cite{Ghosh79}).  If the ratio of magnetic to thermal pressure is
smaller than unity, the disc is then subject to the BH instability,
the internal disc field is amplified and bending waves are damped.
The values of $\alpha$ produced by the BH instability range roughly
from $10^{-3}$ to 0.1 (Hawley et al. \cite{Hawley1}; Brandenburg et
al. \cite{brand2}), the largest values corresponding to the case where
the magnetic field varies on a scale large compared to the disc
semithickness (Hawley \cite{Hawley2}).

\section{Numerical response calculations}
\label{sec:results}

\subsection{Unperturbed magnetic field}
\label{sec:bequil}

For the model calculations presented here, we adopt a radial magnetic
field corresponding to the situation when the central dipole  flux is
completely excluded from the disc.  We consider both the solutions
computed by Aly (\cite{Aly}) and Low (\cite{Low}).  Aly's solution
corresponds to the case when all the dipole flux goes through the
magnetospheric cavity in the middle of the disc, and is singular at
the disc inner edge.  Low gives a solution that is non singular with
some flux escaping to infinity.

In these cases we have, for $r$ larger than some radius $r_B$:

\begin{equation} 
B_r^+ = \mp B_0 \left( \frac{R_o}{r} \right)^3 \left[ \left( \frac{r}{r_B}
\right)^2 -1 \right]^{\pm 1/2} ,
\label{Aly} 
\end{equation}

\noindent where we have defined:

\begin{equation} 
B_0= \frac{4 \mu_d}{\pi R_o^3}.
\label{B0}
\end{equation}  

\noindent In equation~(\ref{Aly}), the upper sign corresponds to Low's
solution, whereas the lower sign corresponds to Aly's solution.  Of
course, in both cases $B_z=0$ since the disc is diamagnetic.  Note
that the radial field at the disc surface has opposite sign in the two
cases.

We comment that $B_r^+$ is negative and positive for Low's and Aly's
cases, respectively, which is the opposite of the solution obtained by
these authors.  This is because we have arbitrarily chosen $B_z$ to be
positive when the dipole moment $\mu_d$ is positive, in contrast to
these authors.

In a two dimensional model, $r_B$ is the disc inner radius.  However,
more realistically it is expected to differ from this by an amount
comparable to the vertical thickness of the disc.  We adopted the
procedure of taking the disc inner radius $R_i$ to be slightly larger
than $r_B$, and to check convergence of the results by decreasing
$\left( R_i - r_B \right)$.  For Aly's solution, convergence requires
that the disc surface density increases rapidly enough towards $R_i$
such as to limit the Lorentz force there.

\subsection{Results}

We have performed global normal mode calculations for disc models with
the unperturbed magnetic field described above.

\noindent We solve equation~(\ref{mod}) considered as an
integro-differential equation for the response function $\xi_z$ by the
method described in APT.  We use the dimensionless radius
$\varpi=r/R_o$, so that the outer radius is $\varpi_o=1$.  In these
units, the disc inner radius on the computational grid is taken to be
$\varpi_i=R_i/R_o=0.1$, and to corotate with the central star so that
$\omega = \Omega_K(R_i)$.

The radial interval $\left[\varpi_i, 1 \right]$ is divided into a grid
of $n_r$ equally spaced points at positions $\left( \varpi_I
\right)_{I=1 \ldots n_r}$ with a spacing ${\Delta \varpi}_I =
\varpi_{I+1}-\varpi_{I}$.  Equation (\ref{mod}) is solved as in APT by
discretizing it on the grid so converting it into a matrix inversion
problem for $\left( \xi_z \right)_{I=1 \ldots n_r}$.  In the
calculations presented below, $n_r$ was varied between 500 and 700,
but 300 already gave satisfactory results.

As already noted in \S~\ref{sec:dispersion}, we use the magnetic
support $\beta$, which would be the ratio of the Lorentz force to the
centrifugal force in the disc if $B_z$ were non zero and on the order
of $B^+_r$:

\begin{equation}
\beta = \frac{ 2 \left( B^+_r \right)^2 r^2}{\mu_0 \Sigma G M_{\ast}} ,
\label{beta1}
\end{equation}

\noindent where $G$ is the gravitational constant.  Although $B_z=0$
in the disc, $\beta$ will be used as a measure of the strength of the
field $B^+_r$.  We found it convenient to define the dimensionless
quantities $\bar{\Sigma} = \Sigma / \Sigma_i$, where $\Sigma_i$ if the
surface density at the disc inner radius, $\bar{B}_r^+ = B_r^+ /B_0$,
and $\bar{\beta} = \beta / \beta_0 $ with:

\begin{equation} 
\beta_0 = \frac{ B_0^2 R_o^2 }{ \mu_0 \Sigma_i G M_{\ast} }.  
\label{beta0}
\end{equation}

\noindent We then have:

\begin{equation}
\bar{\beta} = \frac{2 \varpi^2 \left( \bar{B}_r^+
\right)^2}{\bar{\Sigma}} .
\label{beta}
\end{equation}


For both Low's and Aly's magnetic field, we consider models where the
magnetic support is large (reaching values of order unity) close to
the disc inner edge and decreases rapidly with radius.  For Aly's
solution, we take $\beta=1$ at the disc inner edge.  For Low's
solution, $\beta$ has to vanish at the same location as $B^+_r$,
i.e. at $\varpi_B=r_B/R_o$, otherwise $\xi_z$ is infinite there.  We
then take $\beta=0$ at $\varpi_B$ and $\beta=1$ at the location where
$B^+_r$ has its maximum.  From the value of $\beta$ and $\bar{B}_r^+$
at the disc inner radius, we can calculate $\beta_0$.

From the profile of $\bar{\beta}$, we calculate $\bar{\Sigma}$ using
equation~(\ref{beta}).  We actually take the profile of $\bar{\Sigma}$
which gives the desired $\bar{\beta}$ near the disc inner edge, and we
then match it on to a $r^{-3/2}$ profile further away in the disc
(since $B^+_r$ decreases rapidly with radius, we do not need to worry
about the profile of $\Sigma$ for $\varpi$ larger than a value which
turns out to be about 0.2).

We note that our results do not depend on the numerical values of
$\Sigma_i$, $B_0$, and $R_i$ (or equivalently $R_o$) taken
individually, but only on $\beta_0$ and $\delta$.  For $R_i$ we could
take a few times $R_{\ast}$, and then compute $R_o=10 R_i$.  We could
then get a value of $B_0$ by using equation~(\ref{B0}) with $\mu_d =
B_{\ast} R_{\ast}^3$.  Then $\Sigma_i$ would follow from the
expression~(\ref{beta0}) for $\beta_0$.  Depending on the value of
$\Sigma_i$, the amount of mass in the annulus between $R_i$ and $R_o$
could be modified by changing the profile of $\bar{\Sigma}$ for
$\varpi > 0.2$.

An external poloidal magnetic field tends to squeeze the disc.
Therefore, we choose a form of the aspect ratio $H/r$ which is zero at
the disc inner edge and which increases as the magnetic support
decreases.  The disc is squeezed by the magnetic field if the magnetic
pressure dominates over the thermal pressure.  If the sound speed is
about a tenth of the Keplerian velocity, then the magnetic and thermal
pressures become comparable when the magnetic support is about 0.1.
Therefore, we choose a profile of $H/r$ which reaches a constant value
(that we take to be 0.1) at the location where $\beta=0.1$.

We note that the vertical displacement of the disc at a location $(r,
\varphi)$ is:

\begin{equation}
{\mathcal R}[\xi_z(r)] \cos \varphi - {\mathcal
I}[\xi_z(r)] \sin \varphi.
\label{physxi}
\end{equation}

\begin{figure}
\centerline{
\epsfig{file=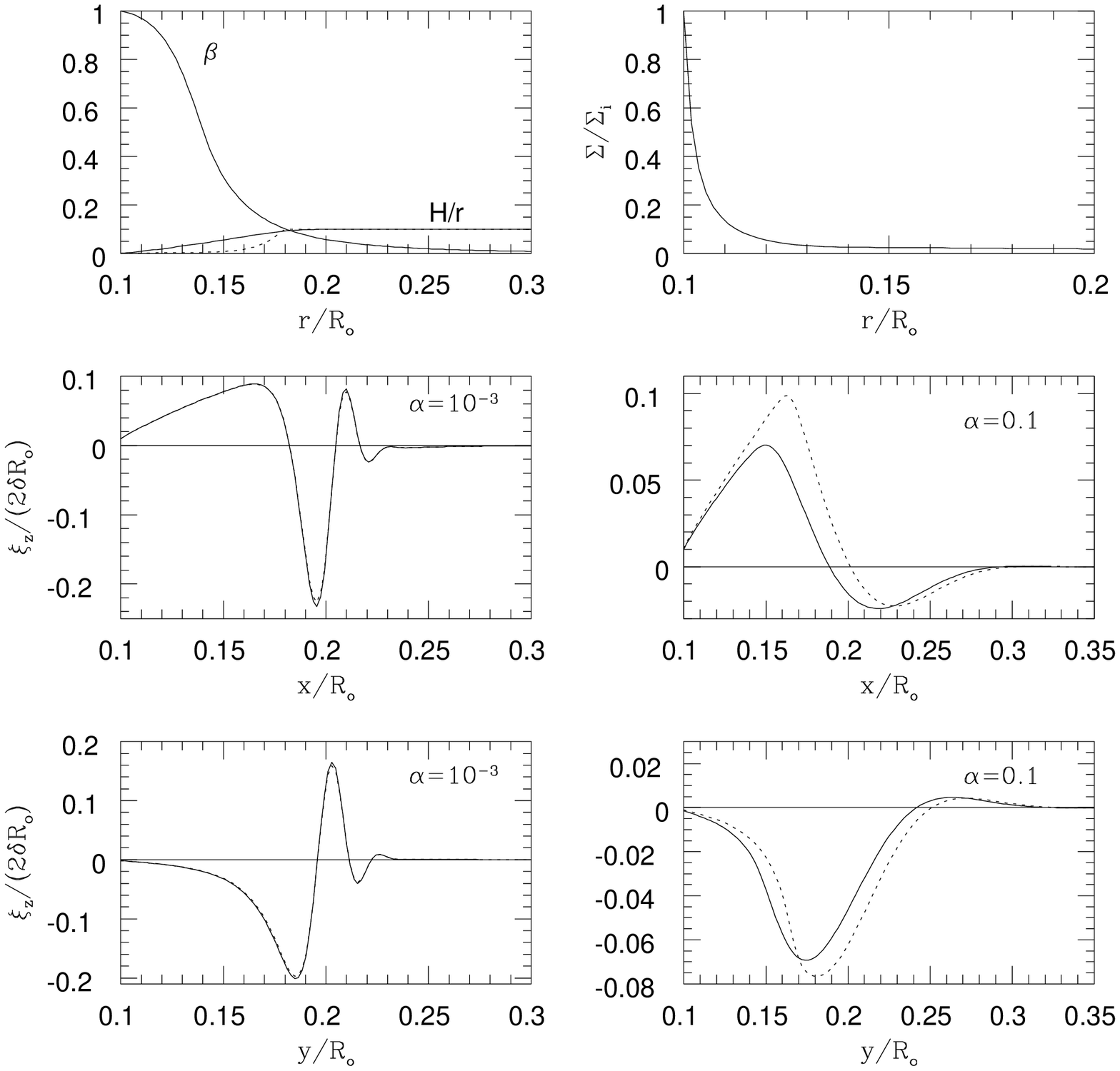,height=16.cm,width=16.cm} }
\caption[]{ Results for Aly's magnetic field.  {\em Top left panel:
\/} Magnetic support $\beta$ and aspect ratio $H/r$ for an almost
linear ({\em solid line}) and exponential ({\em dotted line}) profile
versus $\varpi=r/R_o$.  {\em Top right panel: \/} Surface density
$\Sigma/\Sigma_i$ where $\Sigma_i=\Sigma(R_i)$ versus $\varpi=r/R_o$
{\em Middle panels: \/} ${\mathcal R}(\xi_z)/(2\delta R_o)$, i.e.
$\xi_z/(2\delta R_o)$ for $\varphi=0$ versus $x/R_o$.  {\em Bottom
panels: \/} $-{\mathcal I}(\xi_z)/(2\delta R_o)$, i.e. $\xi_z/(2\delta
R_o)$ for $\varphi=\pi/2$ versus $y/R_o$.  The different curves are
for $\alpha=10^{-3}$ ({\em left panels}) and $0.1$ ({\em right panels
}) and correspond to the almost linear ({\em solid lines}) and
exponential ({\em dotted lines}) profile of $H/r$.  Here
$\varpi_B=0.0975$.  Only the inner parts of the disc where the
displacement is non zero are shown.  Several oscillations
corresponding to bending waves excited at the OLR are visible for
$\alpha=10^{-3}$.  These are damped by viscosity before they propagate
very far.}
\label{fig1}
\end{figure}

Figures~\ref{fig1}--\ref{fig3} show the magnetic support $\beta$, the
disc aspect ratio $H/r$ and the surface density $\Sigma/\Sigma_i$
versus $\varpi$.  Also displayed are ${\mathcal R}(\xi_z)/(2\delta
R_o)$, i.e. $\xi_z/(2\delta R_o)$ for $\varphi=0$ versus $x/R_o$, and
$-{\mathcal I}(\xi_z)/(2\delta R_o)$, i.e. $\xi_z/(2\delta R_o)$ for
$\varphi=\pi/2$ versus $y/R_o$, for $\alpha=10^{-3}$ and 0.1.  These
quantities represent the disc vertical displacement within a factor $2
\delta R_o$ in the $(\varphi=0)$ and $(\varphi=\pi/2)$ half planes,
respectively, or, equivalently, in the $(x>0, z)$ and $(y>0, z)$ half
planes, respectively (see eq.~[\ref{physxi}]).  The different figures
correspond to different models.

Figure~\ref{fig1} and~\ref{fig2} correspond to Aly's solution and
$\varpi_B=0.0975$.  In Figure~\ref{fig1}, $1/\beta \propto {\rm exp}
\left[ 100 (\varpi-0.14) \right] +1$.  The results do not depend
significantly on the details of the $H/r$ profile in the disc inner
parts (we considered both the case where $H/r$ increases almost
linearly and exponentially with radius).  We also checked that $\xi_z$
hardly changes when $\varpi_B$ varies between 0.09 and 0.099,
providing we keep the same magnetic support (i.e. we change $\Sigma$
accordingly).  Figure~\ref{fig2} corresponds to $\beta \propto B^+_r /
\left\{ {\rm exp} \left[ 40 (\varpi-0.14) \right] +1 \right\}$.  The
results are qualitatively the same as in Figure~\ref{fig1}.  We also
ran the case $\beta \propto 1/\varpi^3$, which gave similar results,
except that the positive values reached by $\xi_z$ in that case were
almost as large as the negative values.

Figure~\ref{fig3} corresponds to Low's solution and
$\varpi_B=0.09999999$.  Here again, the results hardly depend on the
details of the profile of $H/r$ in the disc inner parts.  From
equation~(\ref{mod}), we see that $\xi_z$ vanishes at
$\varpi=\varpi_B$ if $\Sigma$ is non zero there.  When $\varpi_B$ is
very close to $\varpi_i$, we indeed check that $\xi_z$ is almost zero
at the disc inner edge.  However, since $B^+_r$ varies very rapidly
near $\varpi_B$, $\xi_z$ takes some finite value at the inner edge as
$\varpi_B$ is moved a little bit away from $\varpi_i$.  This is
illustrated in Figure~\ref{fig4}, where we plot $\beta$ and $\xi_z$
for different values of $\varpi_B$, ranging from 0.09 to 0.09999999
(all the curves have the same parameters except for $\varpi_B$).

The calculations for the low viscosity $\alpha = 10^{-3}$ illustrate
the excitation of bending waves.  Several oscillations corresponding
to bending waves excited at the OLR are visible.  These are damped by
viscosity before they propagate very far.  This damping is enhanced by
the increasing wavenumber produced by the decreasing magnetic field.

For larger and probably more realistic viscosity $\alpha= 0.1,$ these
waves are so heavily damped they are barely visible.  The dominant
displacement is near the disc inner edge in this case.

We note that, when viscous damping is present, the response of the
disc is not in phase with the perturbation, and the warp lags behind
the dipole.  That is, the disc is also twisted, and the twist is
trailing.  Mathematically, this is illustrated by the fact that the
imaginary part of $\xi_z$ is non zero, as seen in the different
figures.  It may seem surprising that, even when $\alpha$ is as low as
$10^{-3}$, the imaginary part of $\xi_z$ is comparable in magnitude to
its real part.  This is because $\xi_z$ varies rapidly with radius
when $\alpha$ is small, and the imaginary part of $\xi_z$ becomes
important when $| \partial^2 \xi_z/\partial r^2 |$ is large (see
eq.~[\ref{mod}]).  We checked however that the imaginary part of
$\xi_z$ becomes very small compared to its real part when $\alpha$ is
reduced further.  Practically, the existence of a twist means that the
point of maximum elevation in the disc is not in the plane which
contains the dipole axis and the rotation axis of the star, i.e. the
$(\varphi=0)$, or equivalently $(x,z)$, plane, but in a plane
corresponding to a smaller (negative) $\varphi$.  Or, in other words,
the elevation does not vanish in the $(\varphi=\pi/2)$, or
equivalently $(y,z)$, plane, but in a plane corresponding to a smaller
value of $\varphi$.  To illustrate this, we show in Fig.~\ref{fig5}
the line of nodes (which joins the points of zero elevation) in the
inner parts of the $(x,y)$ plane for both Aly's and Low's models and
$\alpha=0.1$.  We clearly see that the line of nodes is trailing.  If
there were no twist, it would indeed coincide with the $y$--axis. The
curves corresponding to different values of $\varpi_B$ or different
models are very similar to those displayed in Fig.~\ref{fig5}.  For
$\alpha=10^{-3}$, the line of nodes is more tightly wrapped, because
$\xi_z$ oscillates more rapidly.

In Fig.~\ref{fig6} we present a surface plot of the warped disc
corresponding to the Low's model displayed in Fig.~\ref{fig3}.  This
clearly shows the presence of an obscuring wall like feature near the
disc inner boundary and the trailing twist.

\section{Discussion and application to AA~Tau}
\label{sec:discussion}


The light curve of the classical T~Tauri star AA~Tau displays
photometric, spectroscopic and polarimetric variations on timescales
from a few hours to several weeks (Bouvier et al. \cite{Bouvier2}).
The most striking feature of this light curve is a photometric
variability with a period of 8 to 9 days, comparable to the expected
rotation period of the star.  This has been interpreted by Bouvier et
al. (\cite{Bouvier2}) as being due to the occultation of the star by a
warp of the inner disc (the system is observed almost edge--on).  They
proposed that this warp be produced by the interaction of the disc
with a stellar magnetic dipole tilted with respect to the disc
rotation axis.  Following Mahdavi \& Kenyon (\cite{Mahdavi}), they
assumed that material at the disc inner edge, neglecting warping,
would be more likely to accrete along the shorter field line than
along the longer one connecting star and disc.  In their model, based
on this accretion geometry, the disc is elevated above the original
equatorial plane at the location where the dipole axis is bent toward
the disc.  This geometry enables at least part of the stellar hot
spots, located at the field line footpoints , to still be on the line
of sight when occultation of the star is maximum, in agreement with
the observations.  In this model, the warp is not calculated
self-consistently, and is not actually a real dynamical bending of the
disc.

The assumption by Mahdavi \& Kenyon (\cite{Mahdavi}) can become
incorrect, precisely because it neglects the warping of the disc.  If
the inner disc is no longer in the original equatorial plane, then
material at the inner edge no longer 'sees' the short and long field
lines it would see if it were in the unperturbed disc plane.

\begin{figure}
\centerline{
\epsfig{file=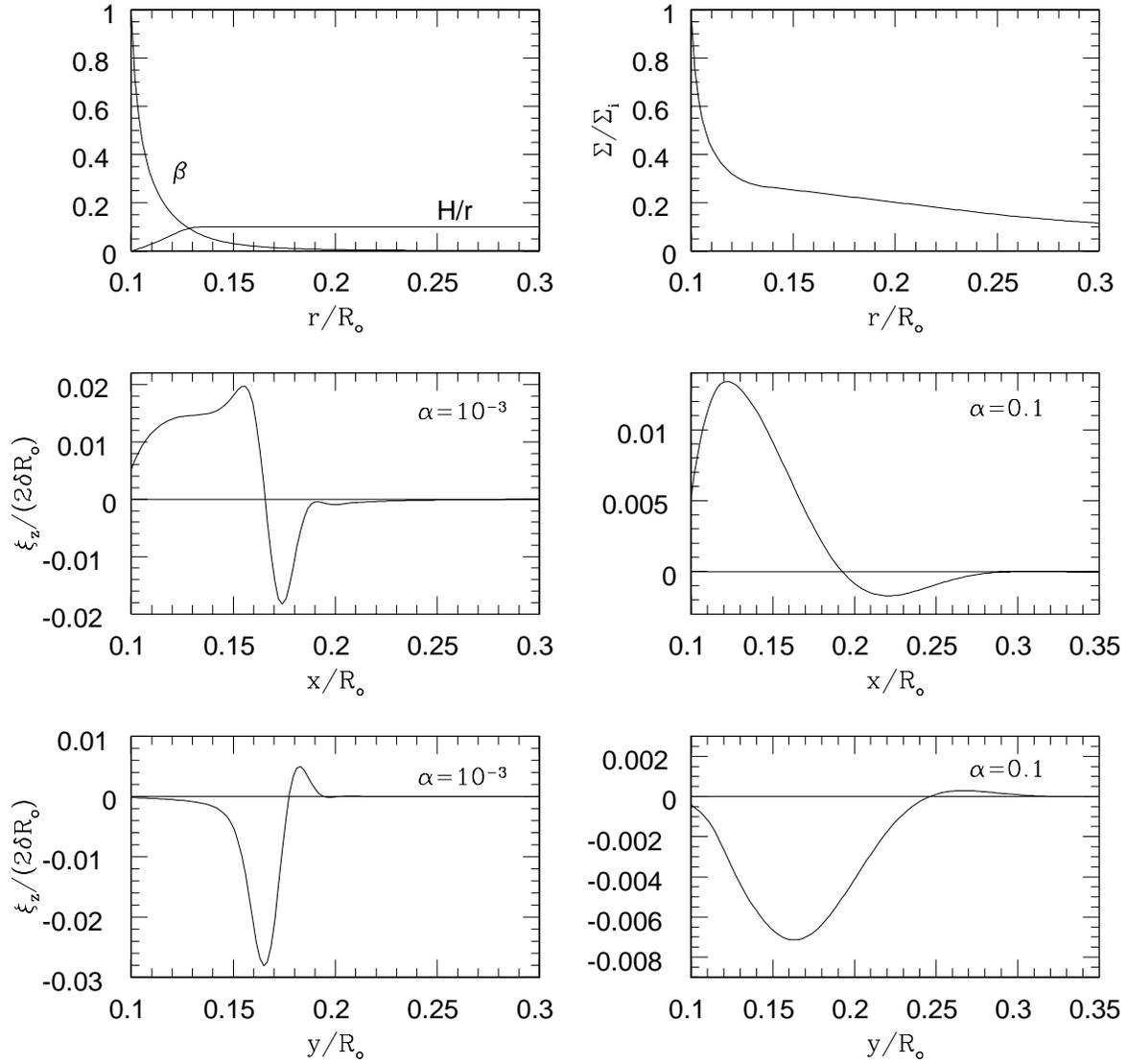,height=16.cm,width=16.cm} }
\caption[]{ Same as Figure~\ref{fig1}, but for a different magnetic
support $\beta$ and surface density $\Sigma$.  Only one profile of
$H/r$ is considered here.  Results are similar, although the maximum
of $|\xi_z|$ is now closer to the disc inner edge.}
\label{fig2}
\end{figure}

\begin{figure}
\centerline{
\epsfig{file=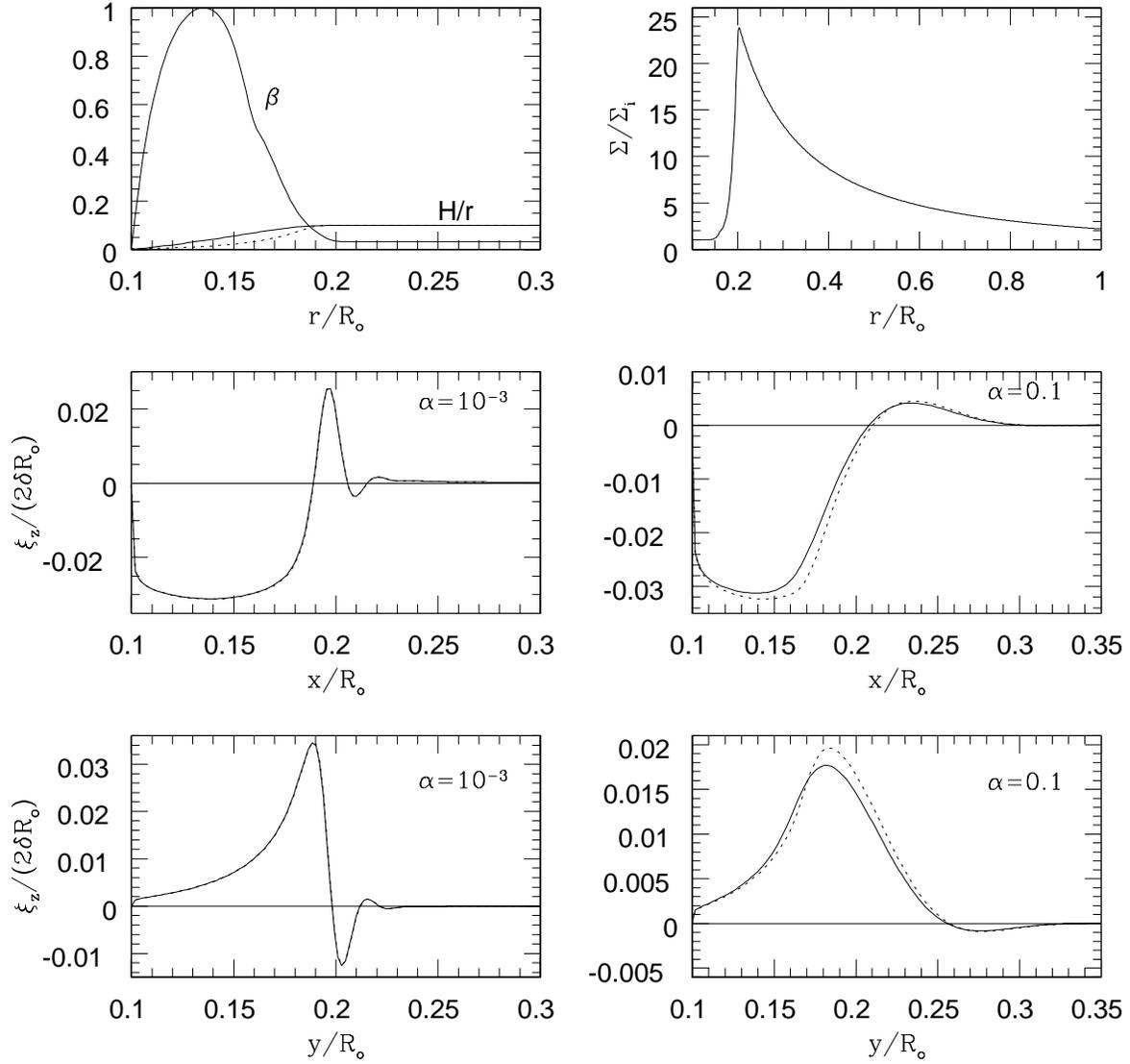,height=16.cm,width=16.cm} }
\caption[]{ Same as Figure~\ref{fig1}, but for Low's magnetic field
with $\varpi_B=0.09999999$ and different magnetic support $\beta$ and
surface density $\Sigma$.  The vertical displacement is almost
constant and maximum very close to the disc inner edge, where it
actually vanishes.}
\label{fig3}
\end{figure}

\begin{figure}
\centerline{
\epsfig{file=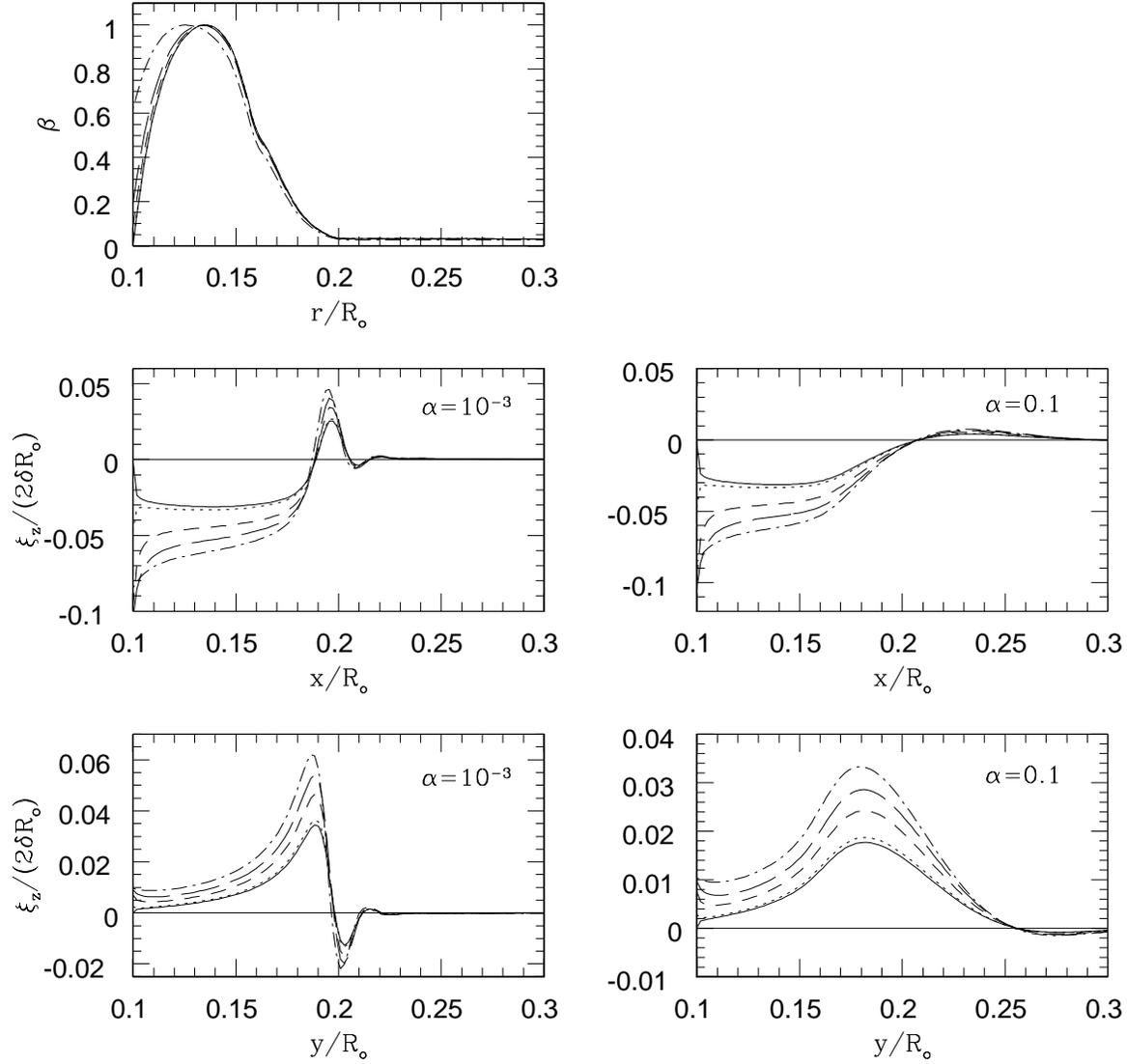,height=16.cm,width=16.cm} }
\caption[]{ Same as Figure~\ref{fig3}, but for different values of
$\varpi_B$.  The different curves correspond to $\varpi_B=0.09999999$
({\em solid lines}), 0.0999 ({\em dotted lines}), 0.099 ({\em
short-dashed lines}), 0.0975 ({\em long-dashed lines}) and 0.09 ({\em
dotted-dahed lines}).  When $\varpi_B$ is moved a little bit further
away from the disc inner edge, the vertical displacement takes a
finite value there.}
\label{fig4}
\end{figure}

\begin{figure}
\centerline{
\epsfig{file=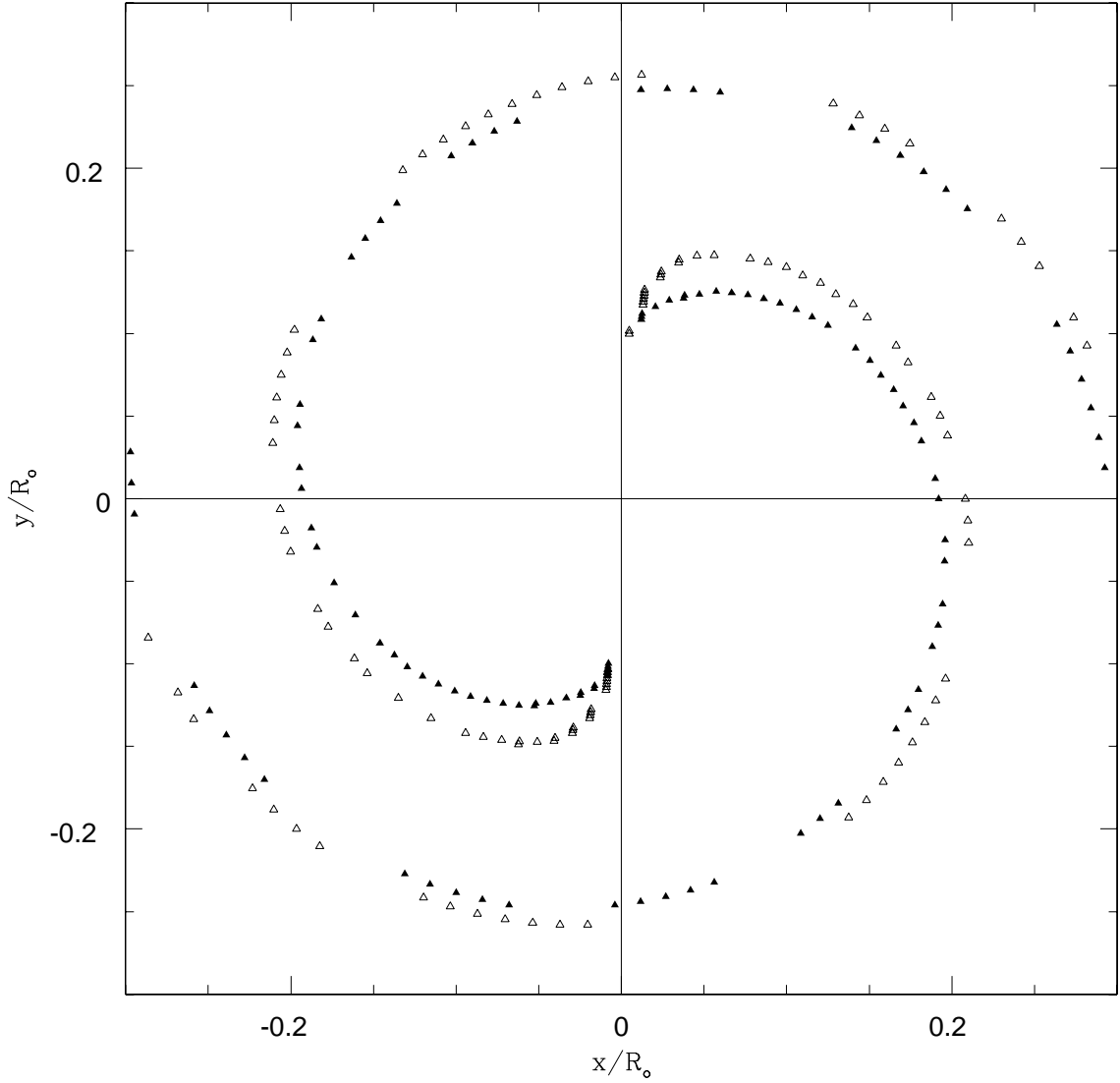,height=16.cm,width=16.cm} }
\caption[]{Line of nodes in the inner parts of the $(x,y)$ plane for
the same models as Figure~\ref{fig2} (Aly's magnetic field, {\em
filled triangles}) and Figure~\ref{fig3} (Low's magnetic field, {\em
open triangles}).  Here the profile of $H/r$ is almost linear and
$\alpha=0.1$. We see that the line of nodes is trailing.  If there
were no twist, it would coincide with the $y$--axis.  For
$\alpha=10^{-3}$, it is more tightly wrapped. }
\label{fig5}
\end{figure}

\begin{figure}
\centerline{
\epsfig{file=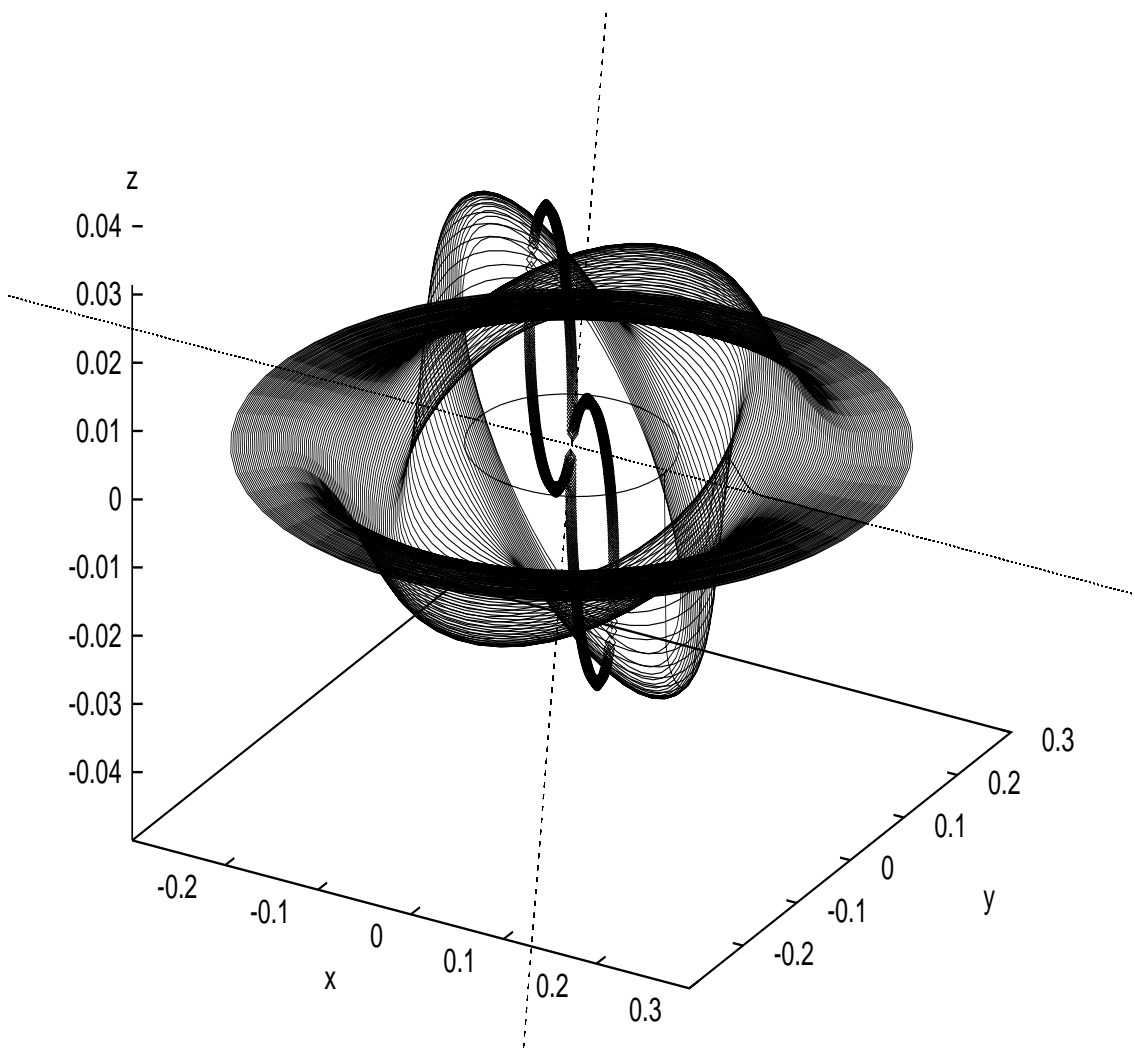,height=16.cm,width=16.cm, angle=270} }
\caption[]{3D view of the disc corresponding to the model represented
in Figure~\ref{fig3} (Low's magnetic field).  The dipole, its axis and
the $x$--axis are also represented.  The vertical scale has been
amplified for clarity.  }
\label{fig6}
\end{figure}

At the disc inner edge, since $\Omega(R_i) = \Omega_{\rm K}(R_i) =
\omega$, we see from equation~(\ref{mod}) that the sign of $\xi_z$ is
the same as that of $P'_m$ (i.e., from eq.~[\ref{Ppert}], the same as
that of $- B_r^{+} B'^{+}_r $) for $\varphi=0$, and that determines
the direction in which the disc bends.  From Figure~\ref{fig3}, we see
that for Low's solution the disc inner edge on the $x$--axis is
pushed below the equatorial plane, which is the opposite of what was
anticipated by Mahdavi \& Kenyon (\cite{Mahdavi}).  For Aly's
solution, the disc is bent above the equatorial plane (see
Fig.~\ref{fig1} and~\ref{fig2}).  It is not surprising that the two
cases give different results since $B_r^+$ changes sign from one case
to another.

\begin{figure}
\centerline{
\epsfig{file=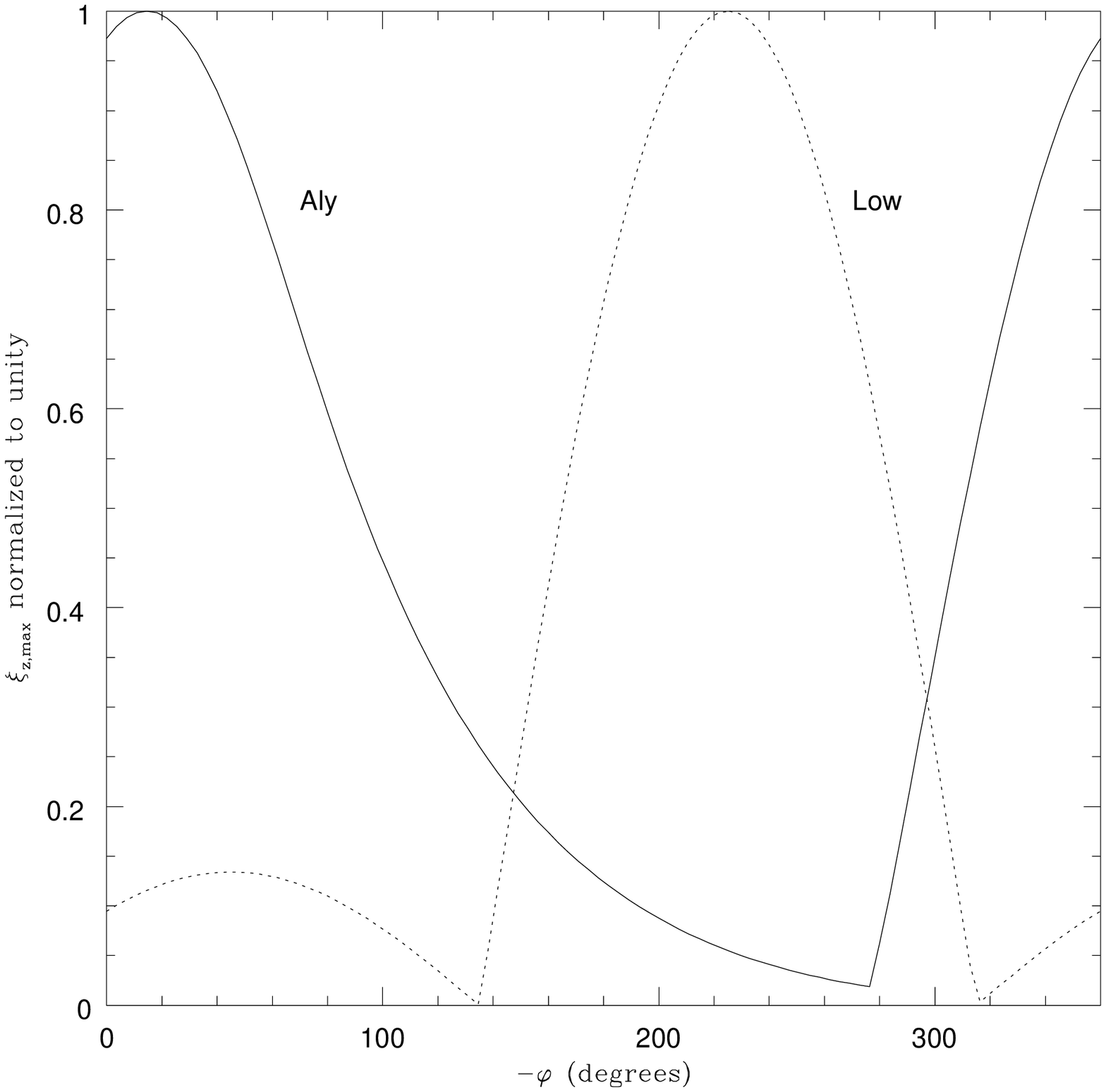,height=16.cm,width=16.cm} }
\caption[]{Maximum elevation $\xi_{z,max}$ (where the $\varphi$
dependence has been taken into account in $\xi_z$), normalized to
unity, above the equatorial plane along the $-\varphi$ direction,
versus $-\varphi$ in degrees.  This represents the maximum elevation
of the disc material located in between the star and an observer
looking at the disc almost edge--on and from above along the
$-\varphi$ direction.  The curves correspond to the models shown in
Figure~\ref{fig2} (Aly's magnetic field, {\em solid line}) and
Figure~\ref{fig3} (Low's magnetic field, {\em dotted line}).  Here the
profile of $H/r$ is almost linear and $\alpha=0.1$.  Depending on the
direction of the line of sight, the radius at which the elevation is
maximum varies (typically between 0.1~$R_o$ and 0.2~$R_o$).  For Aly's
model, maximum occulation occurs for $\varphi$ in the range
$[-\pi/2,0]$, whereas for Low's model it occurs for $\varphi$ in the
range $[-3\pi/2,-\pi]$ or, equivalently, $[\pi/2,\pi]$.}
\label{fig7}
\end{figure}

To get the elevation of the disc above the equatorial plane, the real
and imaginary parts of $\xi_z$ have to be combined through
equation~(\ref{physxi}).  In Fig.~\ref{fig7}, we show the maximum
elevation $\xi_{z,max}$ (where the $\varphi$ dependence has been taken
into account in $\xi_z$), normalized to unity, above the equatorial
plane along the $-\varphi$ direction, versus $-\varphi$.  This
represents the maximum elevation of the disc material located in
between the star and an observer looking at the disc almost edge--on
and from above along the $-\varphi$ direction.  The curves correspond
to Aly's and Low's models displayed in Fig.~\ref{fig1} and~\ref{fig3},
respectively, and $\alpha=0.1$.  Let us suppose that, to begin with,
we look at the system along the $\varphi=0$ direction.  Then, as the
star rotates, our line of sight moves to smaller, negative values of
$\varphi$ (to the right along the $x$--axis of Fig.~\ref{fig7}).  In
the case of Aly's model, this corresponds to the star being occulted
at first and until it has rotated by about 90 degrees.  Then, as the
star rotates further, it becomes less and less obscured, and it can be
seen by the observer until it has completed a full rotation.  In the
case of Low's model, the star is not obscured at first.  Occultation
begins only after the star has rotated by about 180 degrees and lasts
also for about a quarter of a period.  We note that, depending on the
direction of the line of sight, the radius at which the elevation is
maximum varies (typically between 0.1~$R_o$ and 0.2~$R_o$).  It is
unlikely that the observer would be able to know at which radius is
the part of the disc responsible for the occultation, as the whole
structure rotates with the same frequency (that of the star). Indeed,
it is important to stress again that a structure located at some
radius $r$ does not rotate with the local angular frequency but with
that of the star (in other words, the warp appears steady in a frame
rotating with the star).

With Low's solution, the disc is pushed below the equatorial plane at
the location where the dipole is bent toward the disc.  Therefore the
stellar hot spots, at the field line footpoints cannot be seen under
condition of maximum occultation, unless they are relatively large.
In Aly's case, at least part of the hot spots can still be observed at
an occultation maximum because of the phase--lag between the disc
response and the perturbation.  Indeed, we see in Fig.~\ref{fig7}
that, for the particular Aly's model represented there, the
occultation is maximum when $\varphi \sim -15^{\circ}$, and not when
$\varphi=0$.  Therefore, it seems the observations favour Aly's
model.

To fit the observations, Bouvier et al. (\cite{Bouvier2}) needed the
amplitude of the warp to be about 0.3 times the disc inner radius.  We
see from Figures~\ref{fig1}--\ref{fig3} that this is easily attained
here.  There, $\left| \xi_z \right| /(2 \delta R_i)$ ranges from about
0.1 to 1.  For $\delta$ up to about $30^{\circ}$ (for which $\sin
\delta \simeq \delta$ in radians), this gives $\left| \xi_z
\right|/R_i$ from 0.1 to 1.  The lower value could even be made larger
by considering a profile of $H/r$ with an exponential rather than
almost linear increase in Figure~\ref{fig2}.  Therefore only a
moderate misalignment angle is required to produce $\left| \xi_z
\right| /R_i \sim 0.3$.


Figures~\ref{fig1}--\ref{fig3} show that when the viscosity in the
disc is small ($\alpha=10^{-3}$ for instance), the vertical
displacement of the disc varies rapidly, on a scale smaller than the
disc semi--thickness $H$.  In that case, we expect the warp to become
dispersive (Papaloizou \& Lin \cite{Pap1}), and probably the disc to
disrupt.  It is likely that different situations arise when a disc
interacts with a nonaligned dipole, depending on the detailed physics.
In some cases we may get a moderate smoothly varying warp, as
described above, in other cases it may be that the disc breaks up and
reforms.  This may also produce light curve variability.

We note that the bending waves which are excited by the tilted dipole
transport angular momentum (Papaloizou \& Terquem \cite{Pap3}; Terquem
\cite{Terquem1}), which is deposited in the disc if the waves are
damped.  Since the dipole rotates faster than the disc in which the
waves propagate, the resulting torque opposes accretion onto the star.
This effect may produce additional variability, and will be the
subject of another paper.

\section*{Acknowledgements} 

It is a pleasure to thank J\'er\^ome Bouvier for many interesting and
stimulating discussions about AA~Tau and for useful comments on an
earlier draft of this paper, and Claude Bertout for further
information about the variability of CCTS.  We also thank an anonymous
referee and A.~P. Goodson whose suggestions helped to improve the
quality of this paper.  JCBP is grateful to the Laboratoire
d'Astrophysique de l'Observatoire de Grenoble for hospitality and
visitor support.  CT is grateful to the Astronomy Unit at QMW for
hospitality and visitor support through PPARC grant
PPA/V/O/1997/00261.

\newpage

\begin{thebibliography}{}

\bibitem[1997]{Agapitou} 
Agapitou V., Papaloizou J.C.B., Terquem C., 1997, MNRAS 292, 631

\bibitem[1980]{Aly}
Aly J.J., 1980, A\&A 86, 192

\bibitem[1998]{Balbus}
Balbus S.A., Hawley J.F., 1998, Rev. Mod. Phys. 70, 1 

\bibitem[1988]{Bertout}
Bertout C., Basri G., Bouvier J., 1988, ApJ 330, 350 

\bibitem[1993]{Bouvier1}
Bouvier J., Cabrit S., Fernandez M., Mart\'{\i}n E.L., Matthews~J.M., 
1993, A\&A 272, 176

\bibitem[1999]{Bouvier2}
Bouvier J., Chelli A., Allain S., et al., 1999, A\&A 349, 619

\bibitem[1989]{brand1}
Brandenburg A., Tuominen I., Moss D., 1989, Geophys. Astrophys. Fluid
Dyn. 49, 129

\bibitem[1995]{brand2}
Brandenburg A., Nordlund A., Stein R., Torkelsson U., 1995, ApJ
446, 741

\bibitem[1993]{Edwards1} 
Edwards S., Strom S.E., Hartigan P., Strom K.M., Hillenbrand~L.A., 
1993, AJ 106, 372

\bibitem[1994]{Edwards2}
Edwards S., Hartigan P., Ghandour L., Androulis C., 1994,
AJ 108, 1056

\bibitem[1979]{Ghosh79}
Ghosh P., Lamb F.K., 1979, ApJ 232, 259 

\bibitem[1991]{Ghosh91}
Ghosh P., Lamb F.K., 1991, In: Ventura J., Pines D. (eds.)  Neutron
stars: Theory and observation. Dordrecht: Kluwer, p.~363--444

\bibitem[1978]{Goldreich}
Goldreich P., Tremaine S., 1978, Icarus 34, 240

\bibitem[1997]{Goodson}
Goodson A.P., Winglee R.M., Boehm K.-H., 1997, ApJ 489, 199  

\bibitem[1999]{Guenther}
Guenther E.W., Lehmann H., Emerson J.P., Staude J., 1999, A\&A 341, 768 

\bibitem[1994]{Hartmann}
Hartmann L., Hewett R., Calvet N., 1994, ApJ 426, 669

\bibitem[1995]{Hawley1}
Hawley J.F., Gammie C.F., Balbus S.A., 1995, ApJ 440, 742

\bibitem[2000]{Hawley2}
Hawley J.F., 2000, ApJ 528, 462

\bibitem[1996]{Hayashi}
Hayashi M.R., Shibata K., Matsumoto R., 1996, ApJ 468, L37

\bibitem[1999]{Johns} 
Johns-Krull C., Valenti J.A., Hatzes A.P., Kanaan A., 1999, ApJ 501,
L41

\bibitem[1991]{Konigl}
K\"onigl A., 1991, ApJ 370, L39

\bibitem[1999]{Kudoh} 
Kudoh T., Matsumoto R., Shibata K., 1999, In: Nakamoto T. (ed.) Star
Formation 1999. p.~286

\bibitem[1999]{Lai}
Lai D., 1999, ApJ 524, 1030

\bibitem[1986]{Low}
Low B.C., 1986, ApJ 310, 953

\bibitem[1998]{Mahdavi}
Mahdavi A., Kenyon S.J., 1998, ApJ 497, 342

\bibitem[1994]{Mikic}
Mikic Z., Linker J. A., 1994, ApJ, 430, 898

\bibitem[1997]{Miller}
Miller K.A., Stone J.M, 1997, ApJ, 489, 890   

\bibitem[1994]{Montmerle}
Montmerle T., Feigelson E.D., Bouvier J., Andr'e P., 1994, In: Levy
E.~H., Lunine J.~I. (eds.) Protostars and Planets III. Univ. Arizona
Press, Tuscon, p.~689

\bibitem[2000]{Najita}
Najita J., Edwards S., Basri G., Carr J, 2000, In: Mannings V., Boss
A.P., Russell S.S. (eds.) Protostars and Planets IV, Tucson:
University of Arizona Press, in press

\bibitem[1995]{Pap1}
Papaloizou J.C.B., Lin D.N.C., 1995, ARA\&A 33, 505 

\bibitem[1983]{Pap2}
Papaloizou J.C.B., Pringle J.E., 1983, MNRAS 202, 1181

\bibitem[1995]{Pap3}
Papaloizou J.C.B., Terquem C., 1995, MNRAS 227, 553

\bibitem[1973]{Shakura}
Shakura N.I., Sunyaev R.A., 1973, A\&A 24, 337

\bibitem[1984]{Shu} 
Shu F.H., 1984, In: Greenberg R., Brahic A. (eds.) Planetary Rings.
Univ. Arizona Press, Tuscon, p.~513 

\bibitem[1995]{Spruit}
Spruit H.C., Stehle R., Papaloizou J.C.B., 1995, MNRAS 275, 1223 

\bibitem[1990]{Tagger}
Tagger M., Henriksen R.N., Sygnet J.F., Pellat R., 1990, ApJ 353, 654

\bibitem[1987]{Tayler}
Tayler R.J., 1987, MNRAS 227, 553

\bibitem[1998]{Terquem1}
Terquem C.E.J.M.L.J., 1998, ApJ 509, 819

\bibitem[1995]{Wang}
Wang Y.-M., 1995, ApJ 449, L153

\end{thebibliography}
\end{document}